\documentclass[reprint,showpacs,amssymb,amsmath,aps,prb]{revtex4-1}
\setcitestyle{numbers,square}
\usepackage{graphics}
\usepackage{graphicx}
\usepackage{bm}
\usepackage{bbm}
\usepackage{amsmath}
\usepackage{amssymb}
\begin{document}

\title{Pair entanglement in dimerized spin-$s$ chains}
\author{A.\ Boette, R.\ Rossignoli, N.\ Canosa, J.\ M.\ Matera}
\affiliation{IFLP-Departamento de F\'{\i}sica-FCE, 
Universidad Nacional de La Plata, C.C. 67, La Plata (1900), Argentina}
\begin{abstract}
We examine the pair entanglement in the ground state of
finite dimerized spin-$s$ chains interacting through anisotropic $XY$ couplings
immersed in a transverse magnetic field, by means of a self-consistent pair
mean field approximation. The approach, which makes no a priori assumptions on
the pair states, predicts, for sufficiently low coupling between pairs, $2s$
distinct dimerized phases for increasing fields below the pair factorizing
field, separated by spin parity breaking phases. The dimerized phases lead to
approximate magnetization and pair entanglement plateaus, while the parity
breaking phases are characterized by weak pair entanglement but non-negligible
entanglement of the pair with the rest of the system. These predictions are
confirmed by the exact results obtained in finite $s=1$ and $s=3/2$ chains. It
is also shown that for increasing values of the spin $s$, the entanglement of
an isolated pair, as measured by the negativity, rapidly saturates in the
anisotropic $XY$ case  but increases as $s^{1/2}$ in the $XX$ case, reflecting
a distinct single spin entanglement spectrum.
\end{abstract}
%\pacs{03.65.Ud,03.67.Mn,75.10.Jm,64.70.Tg}

\maketitle
\section{Introduction}

The study of entanglement in interacting spin systems has received strong
attention in recent years \cite{Am.08,ECP.10}. Entanglement has provided a
novel perspective for the analysis of correlations and quantum phase
transitions \cite{Am.08,ECP.10,ON.02,VLRK.03}, and is essential for
determining the potential  of such systems in the field of quantum information
\cite{NC.00,HR.06}. Interest in spin systems has  been also enhanced  by the impressive
advances in  control techniques of quantum systems,  that have
made it possible to  simulate interacting spin models with different type of
couplings by means of  trapped ions, Josephson junctions or cold atoms in
optical lattices \cite{PC.04,GA.14,BR.12,SR.15,B.16,LS.12}.

In particular, dimerized systems, characterized by strongly coupled spin pairs
are of great interest, providing a basis for understanding magnetization
plateaus and non-trivial magnetic behavior \cite{ORSB.15}. The phenomenon of 
dimerization can arise from distinct geometric configurations and couplings
\cite{ORSB.15,MG.69,S.05,Pr.75,Pr.77,FM.07,KSV.07,HH.11,KM.08,SR.09,MV.12,Mn.14,HXG.08,GG.09,CRM.10,LM.15,BRCM.15}, 
and has also been recently simulated with cold atoms in optical lattices \cite{ZZ.14}. 
While the basic case deals with singlet pairs in frustrated antiferromagnetic (AFM)-like 
systems \cite{MG.69,LM.15}, other types of dimerization can also arise in the ground state (GS) 
of systems withnon-uniform couplings, like spin chains with alternating-type $XYZ$
couplings \cite{Pr.75,Pr.77,FM.07,HXG.08,GG.09,CRM.10,BRCM.15}. In
these systems a basic mean field approximation (MF) based on independent spins
clearly fails to provide even the most basic features of the GS and its
magnetic behavior. Instead, we have shown \cite{BRCM.15}  that a pair MF
approach, a particular case of a generalized cluster-type variational mean
field treatment, based on independent pairs whose state is self-consistently
determined and admitting relevant symmetry-breaking, is able to provide a
correct basic description. In dimerized spin $1/2$  arrays with
anisotropic $XY$ and $XYZ$ couplings the approach is in fact analytic,
providing a phase diagram that differs from that of the standard  MF and
contains a (single) dimerized phase at low fields under appropriate conditions
\cite{BRCM.15}. Such prediction is in good agreement with the exact results,
which in the special case of spin $1/2$ chains with first neighbor $XY$
couplings can  be analytically obtained through the Jordan-Wigner 
fermionization \cite{Pr.75,Pr.77,CRM.10,LSM.61}.

The aim of this work is to extend previous results to spin $s$ systems with
$s\geq 1$, which are also of interest \cite{KM.08,SR.09,MV.12} and where the previous 
fermionization is no longer available, with the system Hilbert space dimension becoming 
rapidly very large  as the total number of spins increases. In this scenario we
will show that the self-consistent pair MF approach constitutes a convenient method
for understanding the basic physics,  which can still depart considerably from
the conventional MF prediction and the bosonic-like behavior expected for high
spin \cite{MRC.10}. The approach also provides an accurate description of the
reduced state of pairs, enabling to determine   the main  features of the
pair entanglement. In particular, for sufficiently low coupling and appropriate
anisotropies, the approach predicts  $2s$  dimerized phases for increasing
fields below the factorizing field \cite{GG.09,CRM.10,KTM.82}, characterized by
magnetization and pair entanglement approximate plateaus, which are separated
by $S_z$ parity breaking phases where the pair entanglement drops considerably
while that of the pair with the remaining chain becomes non-negligible. These
features are confirmed by the exact numerical results obtained in small finite
spin $1$ and $3/2$ chains. We will also analyze the behavior for large spin,
showing the distinct entanglement properties of anisotropic $XY$ and $XX$
pairs. The formalism and its application to  dimerized spin $s$ $XY$ systems is
described  in sec.\ \ref{II}, while results are discussed in detail in sec.\
\ref{III}.  Conclusions are given in \ref{IV}.

\section{Formalism \label{II}}

\subsection{Pair mean field in dimerized arrays and parity breaking}

We consider a finite chain of $2n$ spins $s$ in a transverse uniform field $B$
interacting through alternating first neighbor anisotropic $XY$ couplings 
\cite{Pr.75,Pr.77,FM.07,GG.09,CRM.10}, such that the chain contains strongly
coupled pairs weakly interacting with their neighboring pairs. The Hamiltonian
can be written as 
\begin{equation}
H=\sum_{i=1}^n [B (S^z_{2i-1}+S^z_{2i})-\sum_{\mu=x,y}J_\mu(S_{2i-1}^\mu
 S_{2i}^\mu+\alpha S_{2i}^\mu S_{2i+1}^\mu)]\,, \label{H1}\end{equation}
where $S_i^\mu$ are the spin components at site $i$ (with
$S_{2n+1}^\mu=S_{1}^\mu$ ($0$) in the cyclic (open) case),  $J_\mu$ are the
exchange couplings and the parameter $\alpha$ indicates the relative strength
of the coupling between pairs ($|\alpha|\leq 1$).  Without loss of generality
we can assume (for even $n$ in the cyclic case) $0\leq \alpha\leq 1$ and $J_x\geq 0$,
as their signs can be changed by local rotations around the $z$ axis, which will 
not alter the spectrum nor the entanglement properties of $H$. We can
 also set  $|J_y|\leq J_x$ by conveniently choosing the $x$ axis and $B\geq 0$ 
(its sign is changed by a global rotation around the $y$ axis). 
The relevant symmetry for $J_y\neq J_x$ is
the $S^z$ parity  
\begin{equation}P_z=\exp[\imath\, \pi\sum_{i=1}^{2n}(S_{i}^z+s)]=
\prod_{i=1}^{2n} P_{zi}\,,\end{equation} 
satisfying $[P_z,H]=0$.  It implies $\langle S^x_i\rangle=\langle S^y_i\rangle=0$ 
$\forall i$ in any non-degenerate eigenstate of $H$. 

In a  pair mean field (MF) treatment, the GS of (\ref{H1}) is approximated by a
pair product state $|\Psi_{0}\rangle=\prod_{i=1}^n|\psi_{0i}\rangle$, with
$|\psi_{0i}\rangle$ the state of  the pair $(2i-1,2i)$.  Minimization of 
$\langle \Psi_0|H|\Psi_0\rangle$ then leads to the independent pair
self-consistent Hamiltonian $h=\sum_{i=1}^n h_i$, with
\begin{eqnarray}
h_i&=&B (S^z_{2i-1}+S^z_{2i})\nonumber\\&&-\sum_{\mu}J_\mu [S_{2i-1}^\mu
 S_{2i}^\mu+\alpha(\langle S_{2i-2}^\mu\rangle S_{2i-1}^\mu+
\langle S_{2i+1}^\mu\rangle S_{2i}^\mu)]\,,\nonumber\\&& \label{h}
\end{eqnarray}
where $\langle S^\mu_{2i+j}\rangle=\langle\psi_{0i}|S^\mu_{2i+j}|\psi_{0i}\rangle$,
$i=1,\ldots,n$, $j=-1,0$, are the mean values in the GS $|\psi_{0i}\rangle$ of
$h_i$: $h_i|\psi_{0i}\rangle=E_{0i}|\psi_{0i}\rangle$  ({\it self-consistency conditions}).  
The essential difference with a conventional MF is that the internal coupling of the pair 
is treated {\it exactly}. Parity breaking is still required for a non-zero average coupling 
between pairs, but the possibility of  parity preserving {\it dimerized} solutions is now open. 
The approximate GS energy is $\langle \Psi_0|H|\Psi_0\rangle=\sum_{i=1}^n[E_{0i}+
\alpha\sum_\mu J_\mu \langle S_{2i}^\mu\rangle\langle S_{2i+1}^\mu\rangle]$, and in case of 
several self-consistent solutions, that with the lowest energy is to be selected. 
 
In the setting considered ($J_x>0$, $|J_y|<J_x$, $\alpha>0$), the pair MF of lowest energy 
is reached for $\langle S^y_{i}\rangle=0$ $\forall$ $i$, since,  writing 
$(\langle S_i^x\rangle,\langle S_i^y\rangle)=|\langle \bm{S}_i^{\perp}\rangle|(\cos\phi_i,\sin\phi_i)$, 
the  MF coupling energy between adjacent pairs, 
$-\alpha\sum_{\mu} J_\mu\langle S^\mu_{2i}\rangle\langle S^\mu_{2i+1}\rangle 
\propto -(J_x\cos\phi_{2i}\cos\phi_{2i+1}+J_y\sin\phi_{2i}\sin\phi_{2i+1})$  
is clearly minimized for $\phi_{2i}=\phi_{2i+1}=0$ (or $\pi$) if $|J_y|<J_x$. 
These values of $\phi_i$ also minimize the internal coupling energy of the pair 
for fixed values of the $|\bm{S}^{\perp}_i|$. We have in fact  verified that parity breaking 
self-consistent pair MF solutions have in the present case either 
$\langle S^x_i\rangle\neq 0$, $\langle S^y_i\rangle=0$ or  $\langle S^x_i\rangle=0$, 
$\langle S^y_i\rangle\neq 0$, but the latter never provides a lower energy. 
In the cyclic case we can also assume for the chosen setting a {\it uniform} pair mean field such that  
$\langle S^x_{2i-1}\rangle=\langle S^x_{2i}\rangle\equiv\langle S^x\rangle$ 
$\forall$ $i$. In the open case the site dependence of $\langle S^x_i\rangle$ 
should be determined self-consistently, but the solution for a uniform $B$ will be practically 
uniform except for small border corrections at the endpoints.   

Hence, the pair MF can be  here characterized by a single parity breaking 
 parameter $\langle S^x\rangle$: If $\langle S^x\rangle=0$ it leads to a 
 parity preserving {\it dimerized phase} at the pair
MF level, with no average coupling  between pairs, while if $\langle
S^x\rangle\neq 0$ it corresponds to  a {\it parity breaking phase}, with
non-zero coupling  between pairs. This last phase is, of course, twofold
degenerate for $|J_y|<J_x$, as both signs $\langle S^x\rangle=\pm|\langle
S^x\rangle|$ are equally possible, with
$|\psi_{0i}^-\rangle=P_{zi}|\psi_{0i}^+\rangle$. At the parity breaking phases
we will then consider the definite parity combinations
\begin{equation}|\Psi_{0\pm}\rangle\propto(\mathbbm{1}\pm P_z)\prod_{i=1}^n|\psi_{0i}^+\rangle
=\prod_{i=1}^n|\psi_{0i}^+\rangle\pm\prod_{i=1}^n|\psi_{0i}^-\rangle\label{pmfr}\,,
\end{equation}
which satisfy $P_z|\Psi_{0\pm}\rangle=\pm|\Psi_{0\pm}\rangle$ and correctly lead to 
$\langle S^x_i\rangle=0$ $\forall$ $i$,  selecting that of
lower energy. Note that these states possess a {\it finite} entanglement between pairs. 

\subsection{Critical conditions}
In order to determine the onset of parity-breaking, we may consider the first
order expansion of the common pair ground state
$|\psi_0\rangle=|\psi_{0i}\rangle$ for small $\langle S^x \rangle$,
 $|\psi_0\rangle \approx |\psi_0^0\rangle + |\delta \psi_0\rangle$,
where $|\delta \psi_0\rangle =\alpha J_x\langle S^x\rangle \sum_{k>0}
\frac{\langle \psi_k^0| S^x_t|\psi_0^0\rangle}{E_k-E_0}|\psi_k^0\rangle$, with
$S^x_t=S^x_{1}+S^x_{2}$ and  $\{|\psi_k^0\rangle\}$ the eigenstates of the
$\langle S^x\rangle=0$ pair Hamiltonian $h^0$:
$h^0|\psi^0_k\rangle=E_k|\psi^0_k\rangle$. Since pair parity symmetry, exactly
conserved in $h^0$, implies $\langle\psi_0^0|S^x_t|\psi_0^0\rangle=0$
(assuming $|\psi_0^0\rangle$ non-degenerate) we have $\langle S^x\rangle\approx 
{\rm Re}[\langle \psi_0^0|S^x_t|\delta\psi_0\rangle]$ up to first order in 
$\langle S^x\rangle$, implying the critical condition
\begin{equation}
1=\alpha J_x\sum_{k>0}\frac{|\langle\psi_k^0|S^x_t|\psi_0^0\rangle|^2}{E_k-E_0}
\,. \label{crit}\end{equation}
 Parity breaking is then feasible if
\begin{equation}\alpha>\alpha_c=\frac{1}{J_x \sum_{k>0}
\frac{|\langle \psi_k^0|S^x_t|\psi_0^0\rangle|^2}{E_k-E_0}}
\,.\label{alphac}\end{equation}
Eq.\ (\ref{alphac}) determines a {\it finite} threshold value $\alpha_c$ 
for parity breaking whenever the isolated pair is {\it gapped} ($E_k-E_0>0$ $\forall$ $k>0$), which
will depend on the relative field strength $B/J_x$, the ratio $\chi=J_y/J_x$
and the spin $s$. The sum in (\ref{crit}) is typically exhausted by the first
term $\frac{|\langle \psi_1^0|S^x_t|\psi_0^0\rangle|^2}{E_1-E_0}$, where $E_1$ 
is the lowest energy of $S^z$ parity opposite to that of $E_0$. 
The restriction $\alpha\leq 1$ sets also an upper bound on $B/J_x$  for 
parity breaking ($B<B^p_c$).

On the other hand, the threshold value $\alpha_c$ in (\ref{alphac})  {\it vanishes} 
when the smallest excitation energy  $E_1-E_0$ becomes zero. Hence, pair GS transitions 
(level crossings) arising for increasing fields will originate parity breaking phases for finite $\alpha>0$ 
{\it even if $\alpha$ is small}, in which the coupling between pairs cannot be treated perturbatively. 
For small $\alpha$ they will emerge  between  {\it distinct} dimerized phases. 
Hence, several onsets (followed by ``deaths'') of parity breaking as the field increases 
can  take place, as will be shown in the next section. For $s=1/2$  there are 
in fact {\it two} parity preserving  phases for increasing fields if $\chi>0$ and  
$\alpha<\frac{1}{2}\chi$ \cite{BRCM.15}, separated by a {\it single} parity 
breaking window $B_{c1}<B<B_{c2}$.

Such multiple dimerized phases are absent for any spin $s$ in the conventional 
single spin MF (full product state approximation), which in this system becomes
equivalent to the MF treatment of a standard chain with uniform coupling of
strength $J_x(1+\alpha)/2$, being independent of $J_y$ if $|J_y|<J_x$. For any
spin $s$ it leads to a single parity breaking phase for $0\leq |B|<B_{c}^{\rm
mf}\equiv J_xs(1+\alpha)$, where $\langle S^x\rangle=\pm s\sin\theta$ with
$\cos\theta=B/B_c^{\rm mf}$. The pair MF phase diagram  will become 
similar to that of the conventional MF for large $\alpha$, although the upper critical  
 field for parity breaking will be  slightly smaller \cite{BRCM.15} (see next section). 

Nevertheless, for $0<J_y<J_x$ there is one point where {\it both} the single
spin and pair mean field treatments exactly coincide and become rigorously {\it exact}
for {\it any} value of the spin $s$  and the number $n$ of spins, i.e., where
the chain GS completely forgets its dimerized structure, which is the {\it
factorizing field} \cite{GG.09,CRM.10,BRCM.15,KTM.82}
 \begin{equation}
 B_s=J_xs(1+\alpha)\sqrt{\chi}\,,
 \;\;\;\chi=J_y/J_x\,.\label{Bs}
 \end{equation}
At this field the chain exhibits a pair of degenerate {\it completely
separable} parity breaking aligned ground states
\begin{equation}|\pm\Theta\rangle=|\pm\theta,\pm\theta,\ldots\rangle\,,
\label{ths}\end{equation}
with $|\pm\theta\rangle=e^{\mp i\theta S^y}|-s\rangle$ a single spin state with
maximum spin forming an angle $\pm\theta$ with the $-z$ axis, with
$\cos\theta=B_s/B_c^{\rm mf}=\sqrt{\chi}$. In a finite chain the factorizing 
field (\ref{Bs}) is actually that where the {\it last} GS parity transition
takes place \cite{RCM.08,RCM.09} (see next section). Accordingly, the
one-sided  left and right limits of  the exact GS at $B_s$ in a finite chain 
will not be given by the product states (\ref{ths}), but rather by the definite 
parity combinations $|\Theta_{\pm}\rangle\propto |\Theta\rangle\pm|-\Theta\rangle$, with
$P_z|\Theta_{\pm}\rangle=\pm|\Theta_{\pm}\rangle$ \cite{RCM.08,RCM.09}, which
will  be correctly predicted by the symmetry-restored pair MF states
(\ref{pmfr}). A GS transition $|\Theta_-\rangle\rightarrow|\Theta_+\rangle$
will then  take place as $B$ crosses $B_s$.

For AFM type couplings ($J_x<0$ and/or $\alpha<0$ in (\ref{H1})), factorizing
and critical values of $B$ and $|\alpha|$  will take exactly the same previous 
values (with $J_x\rightarrow |J_x|$, $\alpha\rightarrow |\alpha|$).  
Just suitable local rotations are to be applied to the corresponding state, 
as previously mentioned. For instance, if $J_x>0$ but $\alpha<0$, they will 
transform the previous uniform pair state into a N\'eel type pair state 
$|\Psi_0\rangle=|\psi_{0}\rangle|\tilde{\psi}_{0}\rangle|\psi_0\rangle\ldots$
with $|\tilde{\psi}_0\rangle=e^{\imath\pi(S_1^z+S_2^z)}|\psi_0\rangle\propto
P_{z}|\psi_0\rangle$ and $\langle S^x\rangle_{2i+j}=(-1)^{i-1} \langle
S^x\rangle$ for $j=0,-1$. These rotations will not affect entanglement measures.

The pair MF approach can of course be also applied to more complex couplings 
and geometries. For instance, if the coupling between pairs $i$ and $i+1$ contains  
 second and third neighbor terms, i.e.\ 
$-\sum_{\mu}J_\mu\sum_{j,l=1,2}\alpha_{jl}S_{2i-2+j}^\mu S_{2i+l}^\mu$ 
we should just replace the $\alpha$-term in (\ref{h}) by 
$\sum_{j,l=1,2}S^\mu_{2i-2+j}(\alpha_{jl}\langle S^\mu_{2i+l}\rangle+\alpha_{lj}
\langle S^{\mu}_{2i-4+l}\rangle)$. If translational symmetry remains unbroken, 
as will occur for  $\alpha_{11}=\alpha_{22}$ and all $\alpha_{jl}\geq 0$, 
previous equations can be directly applied, leading to the same critical 
condition (\ref{crit}) with $\alpha$ replaced by $\sum_{j,l}\alpha_{jl}$. 

\subsection{Entanglement}
The reduced state of a strongly coupled  pair in the exact GS $|\Psi_0\rangle$
of the chain is given by $\rho_{12}={\rm
Tr}_{3,4,\ldots}|\Psi_0\rangle\langle\Psi_0|$. The entanglement of the pair
with the rest of the chain can be measured through the entanglement entropy
$S(\rho_{12})=-{\rm Tr}\rho_{12}\log_2\rho_{12}$, satisfying  $S(\rho_{12})\leq
2\log_2(2s+1)$ for a pair of spins $s$.  On the other hand, its internal
entanglement can be estimated through the negativity (an entanglement monotone
computable for mixed states of any dimension  \cite{VW.02,ZHSL.99})
\begin{equation} N_{12}=({\rm Tr}\,|\rho_{12}^{\rm t_2}|-1)/2\label{N12}\,,\end{equation}
where $\rho_{12}^{\rm t_2}$ denotes the partial transpose of $\rho_{12}$. Eq.\ (\ref{N12})
is just minus the sum of the negative eigenvalues of $\rho_{12}^{\rm t_2}$. If
 $\rho_{12}$ is pure,  Eq.\ (\ref{N12}) becomes a generalized  entanglement entropy,
\begin{equation}N_{12}=[({\rm Tr}\,\sqrt{\rho_1})^2-1]/2
 =\sum_{i<j}\lambda_i^1\lambda^1_j\,,\label{N12p}\end{equation}
where $\rho_1={\rm Tr}_2\,\rho_{12}={\rm Tr}_2\,|\psi_0\rangle\langle\psi_0|$ is the single spin
reduced state and $\lambda^1_i$ its eigenvalues.  Accordingly, Eq.\
(\ref{N12p}) vanishes for $\rho_1$ pure ($|\psi_0\rangle$ separable) and
reaches its maximum for a maximally mixed $\rho_1$  ($|\psi_0\rangle$ maximally
entangled), in which case $N_{12}=s$ for a pair of spins $s$.

At the pair MF level,  $\rho_{12}$ will be pure in the
parity preserving phases. However, in the parity breaking phases $\rho_{12}$ will become
{\it mixed}  if the parity restored states (\ref{pmfr}) are employed. The latter lead
 to a rank $2$ reduced state of the form
\begin{equation}\rho_{12}\approx \frac{1}{2}(|\psi_0^+\rangle\langle\psi_0^+|
+|\psi_0^-\rangle\langle\psi_0^-|)\,,
 \label{pbm}\end{equation}
if the complementary overlap $|\langle \psi_0^+|\psi_0^-\rangle|^{n-1}$
(negligible if $n$ and $\langle S^x\rangle$ are not too small) is discarded,
whose non-zero eigenvalues are just $\lambda_{\pm}=\frac{1}{2}(1\pm|\langle
\psi_{0}^-|\psi_0^+\rangle|)$. Hence, a non-zero entanglement entropy of the
pair with the rest of the chain will arise at the pair MF level within the
parity breaking phases, which will then satisfy $S(\rho_{12})\leq 1$,  with
$S(\rho_{12})\approx 1$  if the overlap $\langle\psi_{0}^-|\psi_0^+\rangle$ is
also negligible.

At the factorizing field (\ref{Bs}), Eq.\ (\ref{pbm}) becomes {\it exact} (if
$\langle -\theta,-\theta|\theta,\theta\rangle^{n-1}=(\cos^{4s}\theta)^{n-1}$ is
neglected), with $|\psi_0^{\pm}\rangle=|\pm\theta,\pm\theta\rangle$ product
states. Consequently, even with symmetry restoration the exact one-sided limits of
$\rho_{12}$ at $B_s$ will become {\it separable}, i.e. a convex combination of
product states \cite{W.89}, leading to $N_{12}=0$ at this point. Nonetheless,
it will remain {\it mixed}, with eigenvalues
$\lambda_{\pm}=\frac{1}{2}(1\pm\cos^{4s}\theta)$, implying a nonzero
entanglement of the pair with the rest of the chain at the left- and right-hand limits
$B\rightarrow B_s^{\pm}$.

\section{Results\label{III}}
\subsection{The spin-1 case}
\subsubsection{The spin 1 pair}
\begin{figure}
\centerline{\scalebox{.8}{\includegraphics*{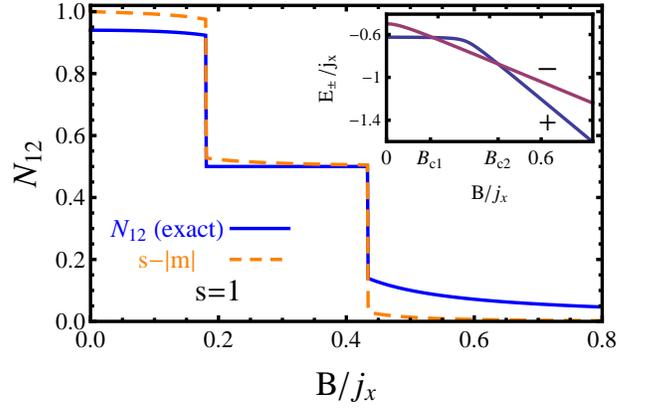}}} 
\vspace*{-0.25cm}

\caption{Negativity of the spin 1 pair GS for an anisotropic $XY$ coupling with
$\chi=J_y/J_x=0.75$, as a function of the scaled transverse  field $B/j_x$,
with $j_x=2J_xs$. The quantity $s-|m|$, with $m$ the intensive magnetization
$\langle S^z_1+S^z_2\rangle/2$, is also depicted. The inset depicts the lowest
energy levels $E_{\pm}$ for each parity, which cross at $B_{c1}$ and
$B_{c2}=B_s$ (pair separability field) and lead to the negativity steps 
(all labels dimensionless in all figures).}
    \label{f1}
\end{figure}

We first examine in detail the case $s=1$. In order to understand the behavior
of both the pair MF and the exact solution for general $\alpha$ in (\ref{H1}),
we first discuss the isolated pair ($\alpha=0$). The lowest energy levels  of
the pair for each parity $P_z$ and for $|J_y|\leq J_x$ are
\begin{eqnarray}
E_+&=&{\textstyle-\sqrt{2B^2+\frac{J_x^2+J_y^2}{2}+\sqrt{4B^2(B^2-J_xJ_y)
+\frac{(J_x^2+J_y^2)^2}{4}}}}\,,\nonumber\\ &&\label{Ep}\\
 E_-&=&{\textstyle-[\frac{J_x+J_y}{2}+\sqrt{B^2 +\frac{(J_x-J_y)^2}{4}}]}\,,
\label{Em}
\end{eqnarray}
with  eigenstates
\begin{eqnarray}
|\psi_+\rangle&=&
\alpha_{-}|-1,-1\rangle+\alpha_0|0,0\rangle+\alpha_+|1,1\rangle+\alpha_{11}
{\textstyle\frac{|-1,1\rangle+|1,-1\rangle}{\sqrt{2}}}\,,\nonumber\\&&
 \label{psip}\\
|\psi_-\rangle&=&{\textstyle\beta_-\frac{|-1,0\rangle+|0,-1\rangle}
{\sqrt{2}}+\beta_+\frac{|0,1\rangle+|1,0\rangle}{\sqrt{2}}
=\frac{|0,\phi\rangle+|\phi,0\rangle}{\sqrt{2}}}\,,\label{psim}
\end{eqnarray}
in the standard product basis $\{|m_1,m_2\rangle\}$ of eigenstates of $S^z_1$
and $S^z_2$, where \begin{eqnarray}
\alpha_0&=&{\textstyle\alpha_-\frac{2(|E_+|-2B)}{J_x-J_y},\;\;\alpha_+
=\alpha_-\frac{|E_+|-2B}{|E_+|+2B},\;\; \alpha_{11}=\alpha_0\frac{J_x+J_y}{\sqrt{2}|E_+|}}
\nonumber\\
\beta_+&=&{\textstyle\beta_-\frac{2(|E_-|-B)-(J_x+J_y)}{J_x-J_y}}\,.
\end{eqnarray}
Here  $|\psi_-\rangle$ is seen to be a Bell type state, with
$|\phi\rangle=\beta_-|-1\rangle+\beta_+|1\rangle$, whereas $|\psi_+\rangle$ has
full Schmidt rank if $J_y\neq J_x$. For strong fields $B\gg J_x$, $E_+\approx
-2B$, $E_-\approx -B$, while for zero field $E_+=-J_x\sqrt{1+\chi^2}$,
$E_-=-J_x$, so that $|\psi_+\rangle$ is the GS in these limits. Yet if
$\chi=J_y/J_x\in(0,1]$,  $|\psi_-\rangle$ will be the GS in an intermediate
field window $B_{c1}\leq B\leq B_{c2}$, as seen in the inset of Fig.\ \ref{f1},
with
\begin{eqnarray}
 B_{c1}&\approx &\sqrt{\chi}J_x{\textstyle\frac{(1-4\chi/25)}{\sqrt{5}}}\label{bc1},\;\;
 B_{c2}=\sqrt{\chi}J_x=B_s\,,\end{eqnarray}
where the expression for $B_{c1}$ holds for small $\chi$ and $B_s$ is the {\it
separability field} (\ref{Bs}) for the isolated pair ($\alpha=0$). Hence, for
$\chi>0$ the pair GS will undergo two parity transitions as the field increases
from 0, the last one at $B_s$.  These transitions are reminiscent of the
magnetization transitions $M\rightarrow M-1$ for $M=0,1$ of the $XX$ case
$J_y=J_x$ ($\chi=1$), where the eigenvalue $M$ of $S^z_t=S^z_1+S^z_2$ is a good
quantum number and $B_{c1}=(\sqrt{2}-1)J_x$,  $B_s=J_x=B^{\rm mf}_c$.
Accordingly, in the $XX$ case the eigenstates (\ref{psip})--(\ref{psim}) become
$|\psi_+\rangle=\frac{1}{\sqrt{2}}(\frac{|-1,1\rangle+|1,-1\rangle}{\sqrt{2}}+|0,0\rangle)$
for $|B|<B_{c1}$ and $|-1,-1\rangle$ for $B>B_s$, with
$|\psi_-\rangle=\frac{|-1,0\rangle+|0,-1\rangle}{\sqrt{2}}$.
Here GS separability holds $\forall$ $B\geq B_s$.

In both the $XY$ and $XX$ cases, these GS transitions lead to  a stepwise
decrease of the pair entanglement, which parallels that of $s-|m|$, with
$m=\langle S^z_t\rangle/2$ the intensive magnetization, as seen in Fig.\
\ref{f1}. Since $|\psi_-\rangle$ is a Bell type state, it has a {\it fixed}
entanglement entropy $S_{12}=1$ and negativity $N_{12}=1/2$, independent of the
anisotropy and field intensity (strict entanglement plateau). On the other
hand, $|\psi_+\rangle$ in (\ref{psip}) leads to a larger negativity for
$|B|<B_{c1}$, not strictly constant, given at zero field by
\begin{equation}N_{12}=
 {\frac{1+|\chi|(1+|\chi|+\chi^2+\sqrt{1+\chi^2})}{2\sqrt{(1+\chi^2)^3}}}\,.
 \end{equation}
This value increases with $|\chi|$ for $|\chi|\leq 1$, reaching
$N_{12}=\frac{1}{4}+\frac{1}{\sqrt{2}}\approx 0.96$ at $\chi=1$ (close to the
maximum value $N_{12}^{\rm max}=1$ for a spin $1$ pair). In contrast, for
strong fields $|B|>B_s$ the state (\ref{psip}) becomes almost aligned, with
just $\alpha_-$ remaining significant, implying a small negativity
$N_{12}\approx \frac{J_x(1-\chi)}{4B}$. As previously stated, 
it is clearly seen that the one-sided limits of the GS at the pair factorizing 
field $B_{c2}=B_s$ are the entangled states $|\psi_{\pm}\rangle$, which at this 
point become linear combinations of the separable states $|\pm \theta,\pm\theta\rangle$ 
(Eq.\ (\ref{ths})).   

The  average magnetization  $\langle S^z_t\rangle$ is given by
$\beta_+^2-\beta_-^2=\frac{-B}{\sqrt{B^2 + J_x^2(1-\chi)^2/4}}$ in
$|\psi_-\rangle$, which is close to $-1$ in the sector where it is GS, and by
$(\alpha_+^2-\alpha_{-}^2)$ in $|\psi_+\rangle$, becoming $\approx
-\frac{B(1-\chi)^2}{J_x(1 +\chi^2)^{3/2}}$ for weak fields $|B|<B_{c_1}$ and
$\approx\frac{J_x^2(1-\chi)^2}{8 B^2}-2$ for strong fields $B>B_s$. Hence, the
behavior of $s-|m|$ resembles that of the negativity,  with $s-|m|\approx
N_{12}^2$ for strong fields.

  \begin{figure}
    \centerline{\scalebox{.88}{\includegraphics*{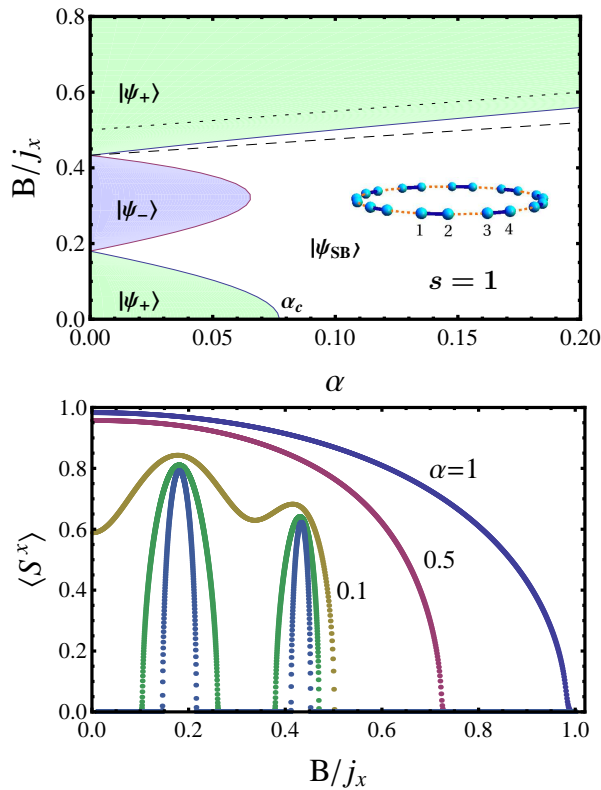}}}
    \vspace*{-0.25cm}
    
\caption{Top: Pair mean field (GMF) phase  diagram of the spin 1
dimerized cyclic chain in the $\alpha$--field plane for $\chi=J_y/J_x=0.75$.
Colored sectors depict dimerized definite $S_z$ parity phases ($\langle
S^x\rangle=0$) whereas the white sector corresponds to the parity breaking phase 
($\alpha_c$ indicates the critical value (\ref{a0}) at zero field). The dashed line depicts
the separability field (\ref{Bs}), entirely contained in the parity breaking phase, which
determines the last parity transition of the exact GS. The dotted line indicates 
the conventional mean field critical field $B_c^{\rm mf}=J_xs(1+\alpha)$.
Bottom: The parity breaking parameter $\langle S^x\rangle$ for increasing
fields at different fixed $\alpha$'s (0.025,0.05,0.1,0.5 and 1). It's behavior
reflects the phases of the top panel, showing a non-monotonous field dependence
at low $\alpha$, with ``deaths and revivals'' if $\alpha<\alpha_c$. For
$\alpha=1$ it lies close to the conventional MF result. }
    \label{f2}
  \end{figure}

\subsubsection{The spin 1 chain}
Returning now to the coupled spin $1$ chain, the previous GS transitions of the
isolated pair will imply three distinct dimerized phases if $\chi>0$ and
$\alpha$ is sufficiently small, as seen in the pair MF phase diagram
depicted in Fig.\ \ref{f2}. For fixed $B<B_c^p\approx J_x s$  Eq.
(\ref{alphac}) determines the threshold value $\alpha_c (B)$ for parity
breaking, which vanishes precisely at the critical fields $B_{c1}$ and $B_{c2}$
of the isolated pair. For $\alpha<\alpha_c(B)$ we then obtain a {\it dimerized
phase} in this approach, with all strongly coupled pairs in a strongly
entangled state $|\psi_+\rangle$ (Eq.\ (\ref{psip})) if $|B|<B_{c1}$  or
$|\psi_-\rangle$  (Eq.\ (\ref{psim}))  if $B_{c1}<B<B_s$, and back to an almost
aligned state $|\psi_+\rangle$ if $B>B_s$. As $B$ increases from $0$ at fixed
small $\alpha$,  the pair MF state can  then undergo {\it four} transitions
between definite parity and parity breaking phases or vice versa, as
shown in Fig.\ \ref{f2}. At zero field, Eq. (\ref{alphac}) leads to the
critical value
\begin{equation}
 \alpha_c(0) = (1 + \chi^2)(\sqrt{1+\chi^2}(4+\chi^2)-4-3\chi^2)/\chi^4\,,\label{a0}
\end{equation}
which increases with $|\chi|$, reaching $\approx 0.14$ for $|\chi|\rightarrow
1$ and vanishing as $\approx \chi^2/8$ for $\chi\rightarrow 0$. For $\chi=0.75$
(Fig.\ \ref{f2}),  $\alpha_c(0)\approx 0.077$.

On the other hand,  if $\alpha>\alpha_c(0)$ we obtain a single parity breaking
phase for $|B|<B_c(\alpha)$, with $B_c(\alpha)$  lying between the factorizing field
(\ref{Bs})  and the standard MF critical field $B_c^{\rm mf}$, as also seen in
Fig.\ \ref{f2}. In the $XX$ limit $\chi=1$, $B_s(\alpha)=B_c(\alpha)=B_c^{\rm
mf}(\alpha)=J_x$.  We also mention that if $\chi<0$ ($-J_x<J_y<0$), the
isolated pair remains gapped for all fields, with a GS which is always of
positive parity and exhibits no sharp transitions. Consequently, in this case
parity breaking occurs just above a {\it finite} threshold
$\alpha_c(B)>\alpha_c(0)$ $\forall$ $B<B_c^p$, i.e., for $\alpha>\alpha_c(0)$
and  $|B|<B_c(\alpha)<B_c^{\rm mf}(\alpha)$. If $\alpha<\alpha_c(0)$ and
$\chi<0$ no parity breaking occurs. Thus, we see that the weaker strength $J_y$
does strongly affect the pair MF phase diagram, in contrast with the single
spin MF.

\begin{figure}
  \begin{center}
 \centerline{\scalebox{.55}{\includegraphics{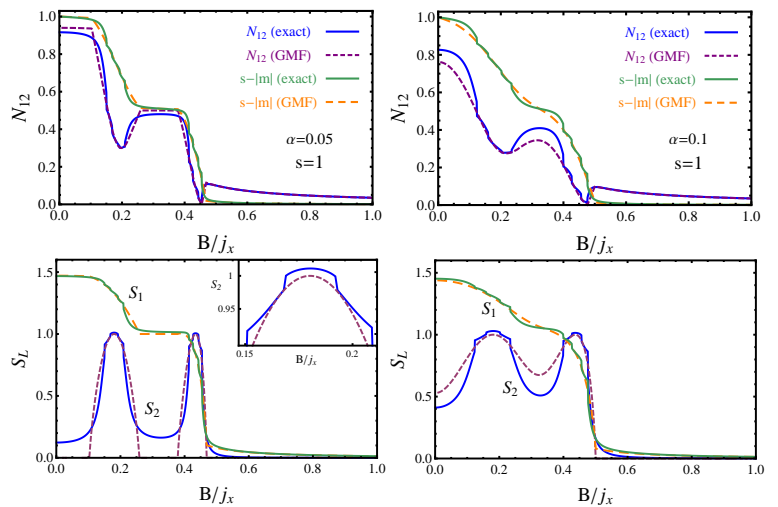}}}
 \vspace*{-0.05cm}
 
\caption{Top panels: Exact and GMF results for the negativity of a strongly coupled pair in
the dimerized spin 1 chain, as a function of the magnetic field, for $\chi=0.75$ and 
coupling factors $\alpha=0.05$ (left) and $\alpha=0.1$ (right). The quantity $s-|m|$,
with $m=\langle \sum_i S_i^z\rangle/2n$ the intensive magnetization, is also depicted.
Bottom panels: The corresponding exact (solid lines) and GMF (dashed lines) results for 
the entanglement entropies of a strongly coupled spin pair ($S_2$) and a single spin 
($S_1$) with the rest of the chain, for the same values of  $\alpha$ and $s$. The
inset shows the discontinuities in the exact $S_2$ stemming from the GS parity
transitions, which occur within the parity breaking phases of the GMF
approach.}
    \label{f3}
  \end{center}
\end{figure}

Fig.\ \ref{f3} depicts in the top left panel results for the exact negativity
of a strongly coupled pair for increasing fields at fixed low $\alpha=0.05$ and
$\chi=0.75$, together with the intensive magnetization $m=\langle \sum_i S_i^z\rangle/(2n)$
 (through the quantity $s-|m|$), obtained numerically by means of diagonalization in a small
cyclic chain of $2n=8$ spins. It is first seen that the  pair MF prediction,
denoted in what follows as GMF (generalized mean field), is in very good
agreement with the exact results. The two dimerized phases for $B<B_s$ lead to
corresponding approximate plateaus in the negativity $N_{12}$ and
magnetization.  In the parity breaking phases,  $N_{12}$  drops considerably,
with  the GMF result remaining accurate if evaluated with the parity restored
mixed state (\ref{pbm}). The vanishing of  $N_{12}$ at the factorizing field
$B_s\approx 0.45 j_x$ is also observed. The behavior of
$s-|m|$, on the other hand, is close to that of the pair negativity but
exhibits just a straight decrease  at the parity breaking sectors,
reflecting actually the behavior of the single spin entanglement entropy $S_1$, 
shown in the bottom panel.

On the right panels we depict results for $\alpha=0.1$, for which the definite
parity dimerized phases are no longer present in GMF yet the order parameter
$\langle S^x\rangle$ still exhibits a non-monotonous evolution with the field
magnitude (Fig.\ \ref{f2}). Accordingly, the exact results still show a
non-monotonous evolution of the negativity, in agreement with the GMF
prediction. The  magnetization plateaus start to disappear, with $m$ again correctly 
predicted by GMF.

The magnetic behavior of the entanglement entropies of a strongly coupled spin
pair ($S_2=S(\rho_{12}))$  and a single spin ($S_1=S(\rho_1)$) with the rest of
the chain are depicted in the bottom panels. That of $S_2$ is quite different
from $S_1$, exhibiting peaks at the GMF parity breaking phases or in general at
the maxima of the GMF parity breaking parameter $\langle S^x\rangle$,
reflecting its  behavior. Parity breaking is then directly indicative of
the entanglement of the pair with the rest of the chain. Dimerization
is also evident through the lower (rather than larger, as in a standard chain)
value of $S_2$ in comparison with $S_1$ for most fields except in the vicinity of 
the factorizing field $B_s$. The behavior
of $S_1$, on the other hand, is qualitatively similar to that of $s-|m|$, since
the latter is here an indicator of the mixedness of the reduced state $\rho_1$
as $\langle S_i^x\rangle=\langle S_i^y\rangle=0$ due to parity symmetry. The
GMF results (obtained with the mixed state (\ref{pbm}) in parity breaking
phases) are again in good agreement with the exact results for both values of
$\alpha$, providing a clear interpretation and correctly predicting the maximum
value $S_2^{\rm max}\approx 1$ in the parity breaking phases. They also yield 
the exact one-sided-limits of these entropies at the factorizing field $B_s$.

\begin{figure}
	\centerline{\scalebox{.7}{\includegraphics{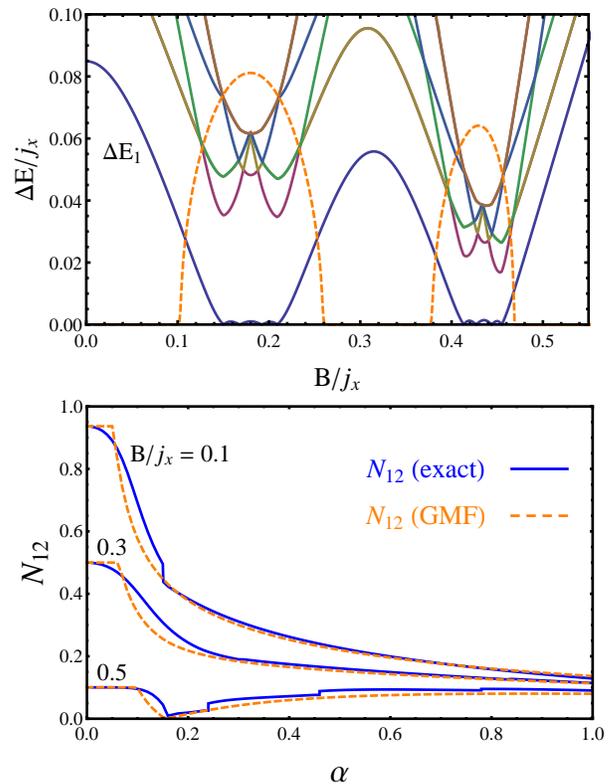}}}
	\vspace*{-0.25cm}
	
	\caption{Top: The lowest exact excitation energies in the $2n=8$
		dimerized spin 1 chain for $\alpha=0.05$ and $\chi=0.75$. The dashed line 
		depicts the (scaled) GMF parity breaking parameter $\langle S^x\rangle$. 
		All parity transitions of the GS are seen to take place within the  GMF parity 
		breaking phases, where the lowest excitation energy becomes small (and the first band 
		of excitation energies minimum). Bottom: The negativity of a strongly coupled pair 
		as a function of the coupling factor $\alpha$ at different fields for the same $\chi$, 
		according to exact and GMF results. }
	\label{f4}
\end{figure}

Moreover,  the  exact  GS of the full chain exhibits $2ns$ parity transitions
as the field increases from $0$ for $\chi>0$, again reminiscent of the $2ns$
magnetization transitions of the $XX$ chain, with the last one  precisely at
the factorizing field  (\ref{Bs}). These transitions {\it are seen to be
confined within the symmetry breaking sectors of the GMF approach}, as
shown in the top panel of Fig.\ \ref{f4} and in the inset of the bottom left panel in Fig.\
\ref{f3} (they lead to small but appreciable discontinuities in all depicted
quantities for small $n$). They indicate the crossings of the lowest negative
and positive parity exact energy levels, which lie very close in the parity
breaking sectors of the pair MF approach, as  verified  in Fig.\ \ref{f4}. 

The pair MF approach remains also quite reliable for larger values of 
$\alpha$, providing very good results for observables such as the pair negativity  
even for $\alpha=1$ (uniform couplings), as seen in the bottom panel of Fig.\ \ref{f4}. 
This result shows that pair MF can improve conventional MF results (which lead to a 
zero negativity at all fields) even for standard (non-dimerized) chains. 
The pair MF transitions at fixed field from the dimerized to the 
parity breaking phase for low increasing $\alpha$, as well as the parity transitions of the exact GS 
can also be observed, together with the vanishing of $N_{12}$ at the factorizing value 
$\alpha\approx 0.15$ for $B/j_x=0.5$. We remark finally that completely similar results are 
obtained in an open chain, with just small corrections for $\langle S^x_i\rangle$ 
at the border pairs. 
\subsection{Higher Spins}

\subsubsection{The spin $s$ case}
\begin{figure}
    \centerline{\scalebox{.9}{\includegraphics*{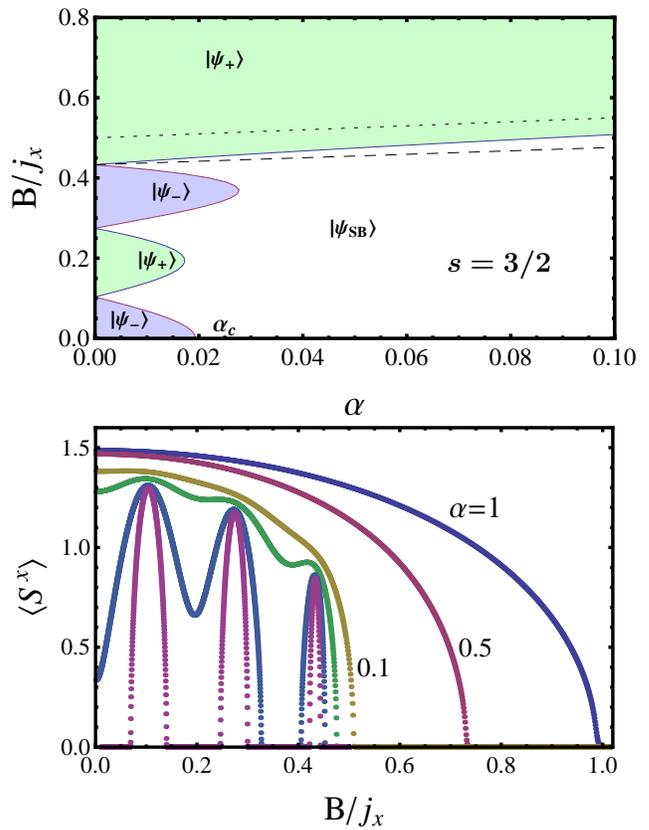}}}
\caption{Top panel: The GMF phase diagram of the spin 3/2
dimerized chain in the $\alpha$-field plane, for anisotropy
$\chi=J_y/J_x=0.75$. There are now three definite parity dimerized phases below
$B_s$ (colored sectors) if $\alpha$ is sufficiently small. Remaining details as
in Fig.\ \ref{f2}. Bottom panel: The corresponding parity breaking parameter
$\langle S^x\rangle$ for increasing fields at different fixed values of
$\alpha$ ($(0.01,0.02,0.05,0.1,0.5$ and $1)$. It's behavior reflects the phases
of the top panel, exhibiting  a non monotonous variation at low $\alpha$, with
``deaths and revivals'' if $\alpha\leq\alpha_c(0)$.}
    \label{f5}
 \end{figure}

The general picture remains similar for higher spins $s$,  but the number of
definite parity dimerized  phases at fixed low $\alpha$ arising for $\chi>0$
and $B<B_s$ in the pair mean field becomes $2 s$, following  the $2s$ GS parity
transitions of the isolated pair for increasing fields (the last one at the
factorizing field for the pair, $B_s=J_xs \sqrt{\chi}$). Consequently, there
are three such phases for $s=3/2$ and $B<B_s$, two of negative parity, as seen in Fig.\
\ref{f5} (for easier comparison between different spins, we scaled the fields
with $j_x=2J_x s$ in all Figures, such that $B_s/j_x$  and $B_c^{\rm
mf}/j_x=(1+\alpha)/2$ are spin independent, with $B_c^{\rm mf}/j_x=1$ in the
uniform ($\alpha=1$) chain). For sufficiently small  $\alpha$ the pair MF GS
can then undergo, for  $s=3/2$, up to six transitions between definite parity and
parity breaking phases (or vice versa) as $B$ increases.

It is also seen that the limit value of $\alpha$ for the existence of multiple
dimerized phases for $\chi>0$ decreases with  increasing spin. At zero field,
we have essentially $\alpha_c(0)\propto\Delta E/(J_x  s^2)$,  with $\Delta
E=E_1-E_0$ the energy gap to the first excited state. For $\chi=1$  ($XX$ case)
$\Delta E\propto J_x$ and hence $\alpha_c(0)\propto s^{-2}$. For $s=3/2$ we
obtain in fact $\alpha_c(0)\approx 0.06$.  However, for $\chi<1$  $\alpha_c(0)$
becomes exponentially small for large $s$, since now $\Delta E$ decreases
exponentially with increasing  spin. The behavior with $s$ of $\alpha_c(B)$ for
other fields $B<B_s$ is qualitatively similar. For $\chi=0.75$  and $s=3/2$ we
obtain $\alpha_c(0)\approx 0.019$, as seen in Fig.\ \ref{f5}.
Nevertheless, for $\alpha$ above but close to $\alpha_c(0)$ the GMF parity
breaking  parameter $\langle S^x\rangle$  continues to exhibit a non-monotonous
evolution for increasing fields, as seen for $\alpha=0.05$, where it still has
three local minima reminiscent of the dimerized phases.  On the other hand, for
$\chi<0$ there is no parity breaking if $\alpha<\alpha_c(0)$, as in the $s=1$
case.

\begin{figure}
    \centerline{\scalebox{.515}{\includegraphics*{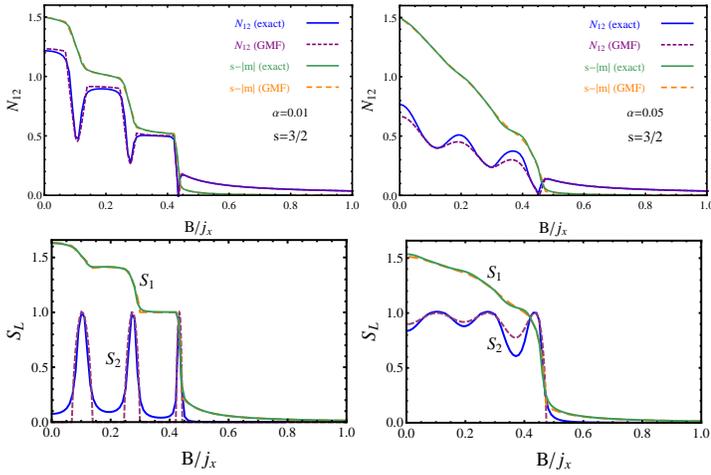}}}
    \vspace*{-0.25cm}
    
\caption{Top: Exact and GMF results for the negativity in the
dimerized spin 3/2 chain,  as a function of the (scaled) magnetic field, for
two different values of the coupling factor $\alpha$. Again $m=\langle
S_z\rangle/n$ denotes  the intensive magnetization. Bottom: The corresponding 
exact (solid lines) and GMF (dashed lines) results for the entanglement
entropies of a strongly coupled spin pair ($S_2$) and a single spin ($S_1$),
with the rest of the chain, as a function of the (scaled) magnetic field in the
$s=3/2$ chain for the previous values of $\alpha$.}
    \label{f6}
\end{figure}

The agreement of the GMF predictions with the exact numerical results remains
high at small values of $\alpha$, as seen in Fig.\ \ref{f6}. The exact
negativity $N_{12}$ and pair entanglement entropy $S_2$ exhibit, accordingly, a
non-monotonous evolution for increasing fields at low $\alpha$, with  $N_{12}$
showing for $\alpha=0.01$ $2s$ approximate plateaus at the GMF dimerized phases
separated by deep valleys at the parity-breaking sectors, before reaching the
strong field regime for $B>B_s$. On the other hand, $S_2$ is again maximum and
close to $1$ at the center of the parity breaking phases, in full agreement
with the GMF result obtained with the parity restored states (\ref{pbm}). We
also see the $2s$ approximate magnetization plateaus, as predicted by GMF.

These effects become  attenuated for $\alpha=0.05$ (right panels), where  the
fully dimerized phases for $B<B_s$ no longer exist in GMF, although  the
behavior of $N_{12}$ and $S_2$ remains non-monotonous, in agreement with
that of $\langle S^x\rangle$ in GMF. It is also seen that the GMF predictions
for the magnetization and the single spin entanglement entropy $S_1$ are very
accurate in both panels, with $s-|m|$ a good  qualitative indicator of the
latter. The exact GS of the finite chain still exhibits $2ns$ parity
transitions as $B$ increases from $0$, the last one at the factorizing field
(\ref{Bs}), although the ensuing discontinuities in the depicted quantities
become small as $s$ increases. They are again confined to the parity breaking
sectors of GMF  (i.e., to the narrow parity breaking intervals for
$\alpha=0.01$). As before, factorization at $B_s$ is reflected in the vanishing
value of $N_{12}$ at this point, while the entanglement entropies $S_1$ and
$S_2$ approach the finite limits determined by the corresponding state
(\ref{pbm}), with $S_2>S_1$ only in the vicinity of $B_s$. 	

\subsubsection{Behavior for large spin}
\begin{figure}
    \centerline{\scalebox{.5}{\includegraphics{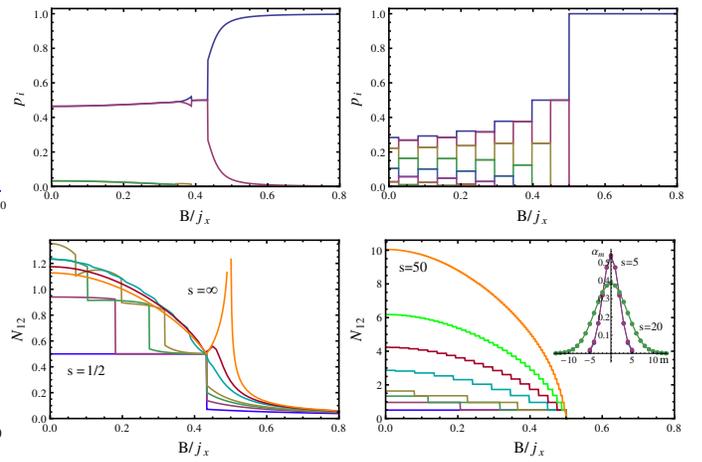}}}
    \vspace*{-0.25cm}
    
\caption{Top: The entanglement spectrum of the GS of a spin $s$
pair as a function of the applied field for $s=5$ and $\chi=J_y/J_x=0.75$
(left) and $1$ (right). In the anisotropic case  it is formed essentially by
just two degenerate eigenvalues below the factorizing field (as described by
Eq.\ (\ref{statesp})), whereas in the $XX$ case (right) there are several
non-vanishing eigenvalues (two-fold degenerate), in agreement with the gaussian
profile (\ref{g}) (shown in the bottom right panel at $B=0$ for $s=5$ and $20$, 
together with the exact results, indistinguishable from (\ref{g})).  
Bottom: The negativity of the pair as a function of the
(scaled) magnetic field for different spin values and $\chi=0.75$ (left), where
$s=1/2,1,\frac{3}{2},2,5,10$ and $\infty$ (bosonic limit, Eq.\ (\ref{N12b})),
and $\chi=1$ (right), where $s=1/2,1,\frac{3}{2},2,5,10,20$ and $50$. Notice
the different scales.}
    \label{f7}
\end{figure}

Let us now examine in more detail the entanglement of a single isolated pair for
increasing spin $s$. In the top panel of Fig.\ \ref{f7} the entanglement
spectrum (the eigenvalues of the single spin reduced density matrix $\rho_1$)
is depicted as a function of the applied field for different spins. For
$\chi<1$ the reduced states become essentially rank 2 states as $s$ increases
in the whole sector $B<B_s$. The reason is that the main component of the pair
GS ($|\psi_{+}\rangle$ or $|\psi_-\rangle$) is just a parity projected rank $2$
mean field state $|\Theta_{\pm}\rangle$, i.e.,
$|\psi_\pm\rangle=\gamma|\Theta_{\pm}\rangle+|\delta\psi_{\pm}\rangle$, with
\begin{eqnarray}
|\Theta_{\pm}\rangle&=&\frac{|\theta,\theta\rangle
\pm|-\theta,-\theta\rangle}{\sqrt{2(1\pm\cos^{4s}\theta)}}=
\sqrt{p_{\pm}}|\theta_+\theta_{\pm}\rangle
+\sqrt{1-p_{\pm}}|\theta_-\theta_{\mp}\rangle  \label{statesp}\,,
\end{eqnarray}
where the last expression is its Schmidt decomposition, with
$|\theta_{\pm}\rangle=\frac{|\theta\rangle\pm|-\theta\rangle}{\sqrt{2(1\pm\cos^{2s}\theta)}}$
the local orthogonal definite parity states and $p_{-}=\frac{1}{2}$,
$p_+=\frac{(1+\cos^{2s}\theta)^2}{2(1+\cos^{4s}\theta)}$. These states lead to a
rank $2$ $\rho_1$ with eigenvalues $(p_{\pm},1-p_{\pm})$. By optimizing the
angle $\theta$ it is found that the overlap
$|\gamma|=|\langle\Theta_{\pm}|\psi_{\pm}\rangle|$ exceeds 0.9 for all field
and spin values, with  $|\langle\Theta_{\pm}|\psi_{\pm}\rangle|\agt 0.95$ for
all fields if $s\geq 5$. The states (\ref{statesp}) become of course the {\it
exact} pair GS at the factorizing field $B_s$, with the overlap staying above
0.99 for $B>B_s$. We can verify from Fig.\ \ref{f7} that the contribution of
$|\delta\psi_{\pm}\rangle$ to the entanglement spectrum is negligible.

Nonetheless, its contribution to the negativity is important if $\chi$ is not
too small. For $|B|<B_s$ the states $|\Theta_{\pm}\rangle$ lead essentially to
an almost constant negativity  $N_{\pm}\approx 1/2$ for large $s$ if $\theta$
is not too small, i.e.,
$N(|\Theta_{+}\rangle)=\frac{1-\cos^{4s}\theta}{2(1+\cos^{4s}\theta)}$,
$N(|\Theta_-\rangle)=\frac{1}{2}$, which for $B<B_s$ lies below the exact
value. The latter remains, however, {\it  bounded as the spin $s$ increases} if
$|\chi|<1$. Its  maximum at zero field is in fact attained at low finite spin
($s\approx 2$ for $\chi=0.75$, as seen in the bottom left panel of Fig.\
\ref{f7}).

For large $s$ and $\chi<1$, the correction $|\delta\psi_{\pm}\rangle$ and its
effect on the pair negativity and entanglement entropy can be determined
through a bosonic RPA (random phase approximation)  approach \cite{MRC.10}. 
Around the normal mean field phase ($B>B_c=J_xs$) such approach implies at lowest 
order the replacements $S_i^z\approx b^\dagger_i b_i-s$, $S_i^{+}\approx 
b^\dagger_{i}$, $S_i^-\approx b_i$ with $b_i$, $b^\dagger_i$ bosonic operators
($[b_i,b^\dagger_j]=\delta_{ij}$), while around the parity-breaking mean field
a similar replacement is to be applied to the rotated spin operators
$S_i^{z'}$, $S_{i}^{\pm'}$, with $S_i^{-'}|\Theta\rangle=0$. Taking into
account parity restoration effects, such bosonisation leads to the analytic
expression
\begin{equation} N_{12}=\left\{\begin{array}{lr}f+\sqrt{f(f+1)}\,,&\;|B|>B_c=J_x s\\
2[f+\sqrt{f(f+1)}]+1/2\,,&\;|B|<B_c\end{array}\right.\label{N12b}\end{equation}
where $f$ is the  average single site bosonic occupation number,
\begin{equation}f={\textstyle\frac{1}{2}(\sqrt{1+\frac{\lambda^2 -\omega_m^2}{\omega_+\omega_-}}-1)}
\,,\end{equation}
with $\lambda=|B|\;(B_c)$ for $|B|>B_c\;(<B_c)$, $\omega_m=\frac{\omega_{+}+\omega_-}{2}$
and $\omega_{\pm}$ the bosonic eigenfrequencies
\begin{equation} \omega_{\pm}=\left\{\begin{array}{lr}
B_c\sqrt{(1\pm(B/B_c)^2)(1 \pm \chi)}\,,&\;|B|<B_c
\\B_c\sqrt{(B/B_c\pm 1)(B/B_c\pm \chi)}\,,&\;|B|>B_c\end{array}\right.\,.\end{equation}
The exact results for the negativity are verified  to approach the previous
finite and $s$-independent bosonic limit for large spin in the anisotropic case
$\chi<1$ (bottom left panel in Fig.\ \ref{f7}). The corresponding pair
entanglement entropy is given by $S_2=-f\log_2 f+(f+1)\log_2(f+1)+\delta$,
where $\delta=0$ ($1$) for $B>B_c$ ($<B_c$) \cite{MRC.10}.

However, in the $XX$ case $\chi=1$ ($J_y=J_x$)  the behavior for high spin is
different. Here $H_{12}$ commutes with the total spin component
$S^z_t=S_1^z+S_2^z$, implying that the parity breaking solution of the pair
mean field is actually breaking a continuous symmetry. Symmetry restoration
implies then integration over all rotations around the $z$ axis (i.e.,
projection onto definite magnetization) and the previous approach (Eqs.\
(\ref{statesp})--(\ref{N12b})) no longer holds. Nevertheless, since the exact
GS has now definite magnetization $M$, it is of the form
\begin{equation}
|\psi_M\rangle=\sum_{m=-s}^{M+s}\alpha_M^m|m,M-m\rangle\,,\;\;\;(\chi=1)\label{sm}
\end{equation}
for $M\leq 0$, with $M$ determined by the applied transverse field ($M\approx -2s [B/B_c]$ for
$B\leq B_c$, $[\ldots]$ integer part) and all $\alpha_M^m$ of the same sign for
$J_x>0$ in (\ref{H1}). Eq.\ (\ref{sm}) is directly its Schmidt decomposition,
implying that the single spin reduced state will have eigenvalues
$|\alpha_M^m|^2$, two-fold degenerate for $m\neq M/2$
($\alpha_M^m=\alpha_M^{M-m}$), leading to the entanglement spectrum of the top
right panel in  Fig.\ \ref{f7}. The number  of non-zero eigenvalues 
(the Schmidt rank of $|\psi_M\rangle$) will then be  $2s+1-|M|$. For $|M|$ not too close 
to $2s$ the coefficients will have essentially a {\it gaussian distribution}, 
as shown in the bottom right panel:  
\begin{equation}\alpha_M^m\propto e^{-(m-M/2)^2/(4\sigma_M^2)}\,,
\;\;\;\sigma_M^2\approx r_M s\label{g}\end{equation}
where for $s$ not too small, the fluctuation $\sigma_M^2\approx
\langle(S^z_1-M/2)^2\rangle$ will be {\it proportional
to the spin $s$}, as obtained from the high spin expansion of the exact eigenvector equation. 
The factor $r_M$ decreases for increasing $|M|$ and for $M=0$ it is given 
by $r_0=1/(2\sqrt{2})\approx 0.35$, whereas for $|M|=s/2$,   
$r_{s/2}\approx 0.32$. The overlap between the gaussian expression
and exact distribution exceeds $0.999$ for $s\geq 5$ at $M=0$. 

Let us remark  that for two spins $s$ coupled to total spin $2s$ and magnetization $M$, 
the corresponding distribution of the  $\alpha_M$'s are  Clebsch-Gordan coefficients, 
which  also lead to a gaussian distribution for high $s$ and $|M|$ not close to $2s$, 
with a fluctuation also proportional to $s$ but slightly smaller 
(at zero field, $\sigma_0^2\approx s/4<s/(2\sqrt{2}))$. Hence, the actual distribution 
in the GS of the $XX$ pair contains small admixtures from lower values of the total spin, 
as $H_{12}$ does not commute with it.  

\begin{figure}
\centerline{\scalebox{.75}{\includegraphics{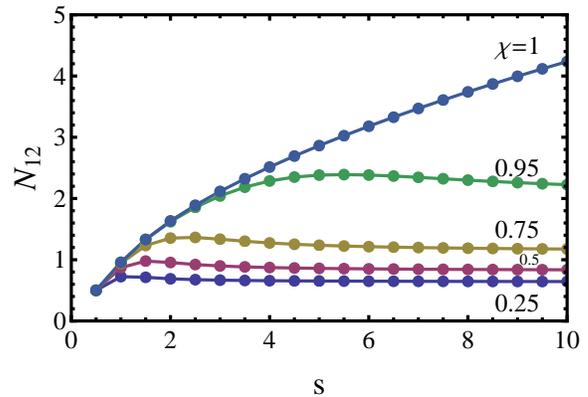}}} 
\vspace*{-0.3cm}

\caption{
Negativity of a pair for increasing spin $s$ at zero field for different
anisotropies $\chi=J_y/J_x$. For $\chi<1$ they saturate, approaching the limit 
values (\ref{N12b}), while for $\chi=1$ they increase as $\sqrt{s}$, as given by 
Eq.\ (\ref{N12a}) (indistinguishable from the exact result for $s\geq 1$ on this scale).}
   \label{f8}
   \vspace*{-0.5cm}
\end{figure}

Therefore, the negativity of the pair can be estimated through the gaussian
approximation, which leads, using Eq.\ (\ref{N12p}), to
\begin{equation}N_{12}\approx \sqrt{2\pi\sigma_M^2}-{\textstyle\frac{1}{2}}\approx
 \sqrt{2\pi r_M s}-{\textstyle\frac{1}{2}}\,.\label{N12a}\end{equation}
Consequently, the entanglement is {\it unbounded} for increasing spin,  with
$N_{12}$  increasing as $\sqrt{s}$ for $\chi=1$, as verified in the bottom
right panel of Fig.\ \ref{f7} and in Fig.\ \ref{f8}.  The  entanglement entropy
of the pair becomes, similarly,  $S(\rho_1)\approx \frac{1}{2\ln 2}[1+\ln(2\pi
\sigma_M^2)]\approx \frac{1}{2\ln 2}[1+\ln(2\pi r_M s)]$.

The previous behavior of the pair entanglement with $s$ holds also for  an
$XXZ$ coupling $-J(S_1^xS_2^x+S_1^yS_2^y)-J_zS_1^zS_2^z$ if  $-J<J_z<J$ ($J>0$),
in which case the coefficients $\alpha_M^m$ remain gaussian with finite width
$\sigma_M$.  However, in the  AFM case $J_z=-J$, $J>0$ (equivalent through
local rotations to $J_z=J<0$) at zero magnetization, the gaussian becomes
uniform and the pair GS becomes {\it maximally entangled}, i.e.
$|\alpha_m^0|=1/\sqrt{2s+1}$ $\forall$ $m$ (with $|\psi_0\rangle$ becoming the
singlet state with zero total angular momentum for $J_z=J<0$). Such state leads
then to $N_{12}=s$ and $S(\rho_1)=\log_2(2s+1)$, with maximum fluctuation
$\langle {S_1^z}^2\rangle=s(s+1)/3$.

\section{Conclusions\label{IV}}
We have shown that  entanglement and magnetization in dimerized spin-$s$ $XY$ chains 
immersed in a transverse field  can exhibit a non-trivial behavior for weak coupling between pairs, 
which lies obviously beyond the scope of a conventional MF description. However, they can be correctly 
described and understood by means of a self-consistent {\it pair} mean field approach. 
Such approach predicts up to $2s$ dimerized phases for increasing 
values of the applied field below the pair factorizing field, if the coupling $\alpha$ 
is sufficiently small and $J_y/J_x>0$, characterized by decreasing values 
of the pair entanglement and lying between parity breaking phases. 
Dimerized sectors are visible in the exact results through approximate plateaus in the  pair 
negativity $N_{12}$ and chain magnetization, and the low values of the
pair entanglement $S_2$ with the rest of the chain, while the intermediate
parity breaking phases through the minima in $N_{12}$ and maxima in $S_2$,
together with  the linear increase in the magnetization.  The latter was also seen to 
correlate with the entanglement entropy $S_1$ of a single spin, which is here larger than 
$S_2$ except in the vicinity of the factorizing field. 

These effects can be all reproduced by the pair MF if symmetry restoration is employed 
in parity breaking phases. These multiple phases arise below increasingly lower values 
of the coupling between pairs as $s$ increases, with the $XX$ case being more favorable. 
Non-monotonous magnetic behavior  of $N_{12}$ and $S_2$ nevertheless persists for higher values, in agreement
with that of the parity breaking parameter of the pair MF. We have also shown that pair MF 
improves conventional MF also for stronger couplings, providing a good prediction 
of the pair negativity (which vanishes identically in conventional MF) for all $\alpha$, including 
the uniform limit $\alpha=1$. 

It was shown as well that the isolated pair GS negativity rapidly saturates as $s$
increases in the anisotropic $XY$ case, in agreement with the predictions of a
MF plus RPA treatment for the pair. However,  in the $XX$ case
it increases as $s^{1/2}$, due to the gaussian-like distribution  (of width $\propto s^{1/2}$) 
of the Schmidt coefficients, being then intermediate between the $XY$ and the full AFM case.

These results indicate that interesting non trivial phases can arise in spin $s$ systems for 
non-homogeneous couplings, which can be predicted by generalized MF approaches  based on 
suitable non-trivial units like pairs or clusters.  While treating the internal couplings exactly, 
they can also account for non-perturbative effects of the coupling between units through symmetry 
breaking. They can in principle be easily implemented in more complex situations (non-uniform fields, 
higher dimensions, etc.), enabling to explore the basic physics of such systems and offering 
a convenient starting point for more involved or specific treatments. 

The authors acknowledge support from CONICET (AB, NC, JMM) and CIC
(RR) of Argentina. Work  supported by CONICET PIP Grant No.\ 112201501-00732.

\end{document}